\documentclass[12pt]{revtex4}

\usepackage{amsmath,amsfonts}
\usepackage{amssymb,latexsym}
\usepackage{color}
\usepackage{graphicx}

\newcommand{\ii}{\mathrm{i}}
\newcommand{\ud}{\mathrm{d}}
\newcommand{\um}{\mathrm{m}}

\newcommand{\uM}{\mathrm{M}}

\begin{document}

\title{Poisson bracket operator}\
\author{T.\ Koide}
\email{tomoikoide@gmail.com,koide@if.ufrj.br}
\affiliation{Instituto de F\'{\i}sica, Universidade Federal do Rio de Janeiro, C.P. 68528,
21941-972, Rio de Janeiro, RJ, Brazil}

\begin{abstract}

We introduce the Poisson bracket operator which is an alternative quantum counterpart of the Poisson bracket. 
This operator is defined using the operator derivative formulated in quantum analysis 
and is equivalent to the Poisson bracket in the classical limit.
Using this, we derive the quantum canonical equation which describes the time evolution of operators.
In the standard applications of quantum mechanics,  
the quantum canonical equation is equivalent to the Heisenberg equation. 
At the same time, this equation is applicable to c-number canonical variables 
and then coincides with the canonical equation in classical mechanics. 
Therefore the Poisson bracket operator enables us to describe classical and quantum behaviors in a unified way.
Moreover, the quantum canonical equation is applicable to non-standard system where the Heisenberg equation 
is not defined. As an example, we consider the application to the system where a c-number and a q-number particles coexist.
The derived dynamics satisfies the Ehrenfest theorem and the energy and momentum conservations.
\end{abstract}
\maketitle 

\section{Introduction}

We reconsider the relation between the Poisson bracket in classical mechanics 
and the commutator in quantum mechanics.
In the canonical quantization,  
the time evolutions of operators are determined by solving the Heisenberg equation which is obtained from the canonical equation by replacing the Poisson bracket
with the commutator,
\begin{eqnarray}
\{ f, g  \}_{PB} = \frac{\partial f}{\partial x} \frac{\partial g}{\partial p} - \frac{\partial f}{\partial p} \frac{\partial g}{\partial x} 
\Longrightarrow 
-\frac{\ii}{\hbar} [f , g] = -\frac{\ii}{\hbar} \left( fg - gf  \right) \,  ,\label{eqn:commu}
\end{eqnarray}
where $f$ and $g$ are functions of canonical variables.
Therefore the commutator is normally considered to be a quantum counterpart of the Poisson bracket.
This identification is however not intuitively understandable. 
For example it is not clear why the derivatives appearing in the Poisson bracket 
are replaced with the non-commutativity of operators.
Moreover the commutator is divided by $\hbar$ in Eq.\ (\ref{eqn:commu}) and hence we cannot see directly 
the classical limit \footnote{ 
Throughout this paper, the classical limit means the disappearance of all commutators. 
This disappearance, however, will not be equivalent to the condition to observe a classical behavior in quantum mechanics.  
That is, the disappearance of all commutators is not a necessary and sufficient condition for the classical limit.
}.

To clarify the role of the Poisson bracket in quantum mechanics, 
we often map operators in the Hilbert space into functions of phase space variables using the Wigner-Weyl transformation. 
In this approach, we can define the Wigner function which is the quasi-probability distribution in the phase space.
The time evolution of the Wigner function is characterized by the Moyal bracket, which is reduced to the Poisson bracket in the classical limit. Thus the Moyal bracket is regarded as a quantum counterpart of the Poisson bracket in this approach.
This perspective is extended in the deformation quantization \cite{dq,bla2013}.

In this paper, however, we do not consider the phase space representation of quantum mechanics.
Instead we introduce the Poisson bracket operator as another quantum counterpart of the Poisson bracket.
This operator is defined through the operator derivative formulated in quantum analysis proposed by Suzuki \cite{qa1,qa2,qa3,qa4,qa5,qa6,qa7}. 
One of the advantages of our approach is that the Poisson bracket operator has 
a clear classical correspondence with the Poisson bracket because 
the operator derivative is equivalent to the standard derivative in the classical limit.
Using this operator, we derive the quantum canonical equation which describes the time evolution of operators.
In the standard applications of quantum mechanics,  
the quantum canonical equation is equivalent to the Heisenberg equation. 
At the same time, this equation is applicable to c-number canonical variables 
and then coincides with the canonical equation in classical mechanics. 
Therefore classical and quantum behaviors are described in a unified way by introducing the Poisson bracket operator 
and this is another advantage.
Moreover, the quantum canonical equation is applicable to non-standard system where the Heisenberg equation 
is not defined. As an example, we consider the application to the system 
where a c-number and a q-number particles coexist.
The derived dynamics satisfies the Ehrenfest theorem and the energy and momentum conservations.

This paper is organized as follows.
In Sec. \ref{sec:qa}, the operator derivative is introduced. 
Three mathematical formulae are introduced in Sec.\ \ref{sec:formulae}. These are used to show the relation between the Poisson bracket operator and the commutator.
The Poisson bracket operator and the quantum canonical equation are introduced in Secs.\ \ref{sec;poisson_op} 
and \ref{sec:qc_eq}, respectively. 
The non-standard application of the quantum canonical equation is discussed in Sec.\ \ref{sec:app}. 
Section \ref{se:conclu} is devoted to concluding remarks.

\section{definition of operator derivative}\label{sec:qa}

We discuss the definition of the operator derivative proposed in quantum analysis \cite{qa1,qa2,qa3,qa4,qa5,qa6,qa7}. 
For other applications of quantum analysis, see, for example, Refs.\ \cite{maj,qse_koide,qse_koide2,ishihara,ciaglia,castro}.

Let us consider an operator $\widehat{A}$ and a function of the operator $f(\widehat{A})$ where $f(x)$ is a smooth function of $x$. Then 
the G\^{a}teaux differential is defined by 
\begin{eqnarray}
\ud
f (\widehat{A}\, ;\widehat{B}) 
= \lim_{h\rightarrow 0} \frac{f(\widehat{A} + h \widehat{B} ) - f(\widehat{A})}{h} \, ,
\end{eqnarray}
where $h$ is a c-number and $\widehat{B}$ is an operator which
is in general not commutative with $\widehat{A}$ \cite{gateaux}.
The operator derivative with respect to $\widehat{A}$ is denoted by ($\ud f/\ud \widehat{A}$) 
and then the G\^{a}teaux differential is expressed as  
\begin{eqnarray}
\ud f(\widehat{A}\, ;\widehat{B}) = \frac{\ud f}{\ud\widehat{A}}\{\widehat{B}\} \, .
\end{eqnarray}
One can see that the operator derivative is a hyper-operator which is operated to $\widehat{B}$. 
In quantum analysis, this operator derivative is defined by 
\begin{eqnarray}
\frac{\ud f}{\ud \widehat{A}}\{\widehat{B}\} = \int^1_0 \ud \lambda \, f^{(1)}(\widehat{A} - \lambda \delta_{\widehat{A}})\widehat{B} \, ,
\label{eqn:def_op_der}
\end{eqnarray}
where $f^{(n)}(x) =\ud^n f(x)/\ud x^n$ and 
\begin{eqnarray}
\delta_{\widehat{A}} = [\widehat{A},\, \, \,\,] \, .
\end{eqnarray}
When an operator is given by a function of time, the time derivative is expressed as  
\begin{eqnarray}
\frac{\ud f(\hat{A}(t))}{\ud t} = \frac{\ud f}{\ud \hat{A}(t)} \left\{ \frac{\ud \hat{A}(t)}{\ud t} \right\} \, .
\label{eqn:def_tderi}
\end{eqnarray}

The above definition can be extended to define the operator partial derivatives. 
Let us consider a smooth function of operators $\widehat{A}$ and $\widehat{B}$ which 
can be expanded as
\begin{eqnarray}
f(\widehat{A},\widehat{B}) = \sum_{n,m \ge 0} f_{nm} \widehat{A}^{n} \widehat{B}^m \, , \label{eqn:op_expa}
\end{eqnarray}
where $n$ and $m$ are non-negative integers and $f_{nm}$ are expansion coefficients.
The partial derivatives with respect to $\widehat{A}$ and $\widehat{B}$ are then defined by 
\begin{eqnarray}
\begin{split}
\frac{\partial f(\widehat{A},\widehat{B})}{\partial \widehat{A}} \{\widehat{C}\} 
&= \sum_{n,m \ge 0} n\, f_{nm} \left( \int^1_0 \ud \lambda \, (\hat{A}-\lambda \delta_{\widehat{A}})^{n-1} \hat{C} \right)
\widehat{B}^m \, , \\ 
\frac{\partial f(\widehat{A},\widehat{B})}{\partial \widehat{B}} \{\widehat{C}\} 
&= \sum_{n,m \ge 0} m\, f_{nm} \widehat{A}^{n} 
\left( \int^1_0 \ud \lambda \, (\widehat{B}-\lambda \delta_{\widehat{B}})^{m-1} \widehat{C} \right) \, ,
\end{split}
\label{eqn:def_par_deri}
\end{eqnarray}
respectively.
When $\widehat{C}$ is given by a c-number, say $c$, the operator derivatives are equivalent to the standard derivatives with respect to c-numbers because $\delta_{\widehat{A}}\, c$ and $\delta_{\widehat{B}}\, c$ vanish, 
\begin{eqnarray}
\begin{split}
\frac{\partial f(\widehat{A},\widehat{B})}{\partial \widehat{A}} \{ c \} &= \sum_{n,m \ge 0} n\, f_{nm} \hat{A}^{n-1} \widehat{B}^m c\, , \\ 
\frac{\partial f(\widehat{A},\widehat{B})}{\partial \widehat{B}} \{ c \} &=  \sum_{n,m \ge 0} m\, f_{nm} \widehat{A}^{n} 
 \widehat{B}^{m-1} c \, .
\end{split}
\end{eqnarray} 
Therefore, in the following calculations, we omit the argument $\{ \,\,\,\, \}$ 
when the operator derivative is operated to a c-number,
\begin{eqnarray}
\begin{split}
\frac{\partial f(\widehat{A},\widehat{B})}{\partial \widehat{A}} 
&= 
\frac{\partial f(\widehat{A},\widehat{B})}{\partial \widehat{A}} \{1\}
\, , \\ 
\frac{\partial f(\widehat{A},\widehat{B})}{\partial \widehat{B}} 
&= 
\frac{\partial f(\widehat{A},\widehat{B})}{\partial \widehat{B}} 
\{1\}
\, .
\end{split}
\end{eqnarray}

Note that the powers of the operators $\widehat{A}$ are ordered to the left of that of $\widehat{B}$ in the expansion (\ref{eqn:op_expa}).
Such a re-ordering of operators is not applicable to general non-commutative operators
and then Eq.\ (\ref{eqn:def_par_deri}) is modified.
For example, the derivative of $\widehat{A}^l\widehat{B}^m\widehat{A}^n\widehat{B}^o$ is given by 
\begin{eqnarray}
\frac{\partial \widehat{A}^l\widehat{B}^m\widehat{A}^n\widehat{B}^o}{\partial \widehat{A}} \{\widehat{C}\} 
&=& \left( l \int^1_0 \ud \lambda \, (\widehat{A}-\lambda \delta_{\widehat{A}})^{l-1} \widehat{C} \right)
\widehat{B}^m\widehat{A}^n\widehat{B}^o \nonumber \\
&& + \widehat{A}^l\widehat{B}^m\left( n \int^1_0 \ud \lambda \, (\widehat{A}-\lambda \delta_{\widehat{A}})^{n-1} \widehat{C} \right)\widehat{B}^o \, ,
\end{eqnarray}
where $n,l,m$ and $o$ are non-negative integers.
In this work, however, we consider exclusively the case where the canonical operators $\widehat{A}$ and $\widehat{B}$ satisfy the commutation relation (\ref{eqn:abcom}).  
Then the expansion (\ref{eqn:op_expa}) is applicable to express arbitrary smooth functions of operators.

\section{Mathematical formulae} \label{sec:formulae}

In this section, we discuss three formulae which are used to show the relation between the Poisson bracket operator and the commutator.

\vspace{1cm}

{\bf Formula 1} 

For arbitrary two operators $\widehat{A}$ and $\widehat{B}$, and an integer $n \ge 1$, 
there exists the following relation, 
\begin{eqnarray}
\int^1_0 \ud \lambda \, (\widehat{A}-\lambda \delta_{\widehat{A}} )^{n} \widehat{B} 
&=&
\frac{1}{n+1} (\widehat{A}^n \widehat{B} + \widehat{A}^{n-1}\widehat{B}\widehat{A} + \cdots + \widehat{B} \widehat{A}^n) \, . \label{eqn:mformula1}
\end{eqnarray}
The proof of this formula is summarized in Appendix \ref{app:formula1}.

The formula (\ref{eqn:mformula1}) is satisfied for any operators.
The other two formulae are however applicable to operators which satisfy a special commutation relation.

{\bf Formula 2} 

Let us consider two operators which satisfies the commutation relation,  
\begin{eqnarray}
[ \widehat{A} , \widehat{B} ] = c \, , \label{eqn:abcom}
\end{eqnarray}
where $c$ is a c-number.
Then, for arbitrary integers $n,m \ge 1$, the following relation is satisfied, 
\begin{eqnarray}
\widehat{B}^n \widehat{A}^m  - \widehat{A}^{m}  \widehat{B}^{n}  
&=& m n (\delta_{\widehat{B}} \widehat{A})  \int^1_0 \ud \lambda \,(\widehat{B} - \lambda \delta_{\widehat{B}})^{n-1} \widehat{A}^{m-1} 
\label{eqn:mformula3}
\, .
\end{eqnarray}
The proof of this formula is summarized in Appendix \ref{app:formula2}.

Using the formula (\ref{eqn:mformula3}), we can show a kind of commutativity associated with the operator derivative.

{\bf Formula 3} 

Let us consider two operators which satisfy the commutation relation,  
\begin{eqnarray}
[ \widehat{A} , \widehat{B} ] = c \, , \label{eqn:abcom}
\end{eqnarray}
where $c$ is a c-number.
Then, for arbitrary integers $n,m \ge 1$, the following relation is satisfied,
\begin{eqnarray}
\int^1_0 \ud \lambda \,(\widehat{B} - \lambda \delta_{\widehat{B}})^{n-1} \widehat{A}^{m-1} 
= \int^1_0 \ud \lambda\, (\widehat{A} - \lambda \delta_{\widehat{A}})^{m-1} \widehat{B}^{n-1} \, .
\label{eqn:mformula2}
\end{eqnarray}
The proof of this formula is summarized in Appendix \ref{app:formula3}.

\section{Poisson bracket operator} \label{sec;poisson_op}

We consider a pair of canonical variables $(\widehat{A},\widehat{B})$ and two smooth functions of these operators,  
$f(\widehat{A},\widehat{B})$ and $g(\widehat{A},\widehat{B})$. 
The Poisson bracket operator is then defined by 
\begin{eqnarray}
\left\{ f(\widehat{A},\widehat{B}), g(\widehat{A},\widehat{B}) \right\}_{(\widehat{A},\widehat{B})} 
\equiv \frac{\partial f(\widehat{A},\widehat{B})}{\partial \widehat{A}}\left\{ \frac{\partial g(\widehat{A},\widehat{B})}{\partial \widehat{B}}  \right\}
- \frac{\partial f(\widehat{A},\widehat{B})}{\partial \widehat{B}}\left\{ \frac{\partial g(\widehat{A},\widehat{B})}{\partial \widehat{A}} \right\}
\, . \label{eqn:def_pb_op}
\end{eqnarray}
From the definition (\ref{eqn:def_op_der}), 
the operator derivative agrees with the standard derivative of c-number 
when the canonical variables $(\widehat{A},\widehat{B})$ are commutative. 
Therefore the classical limit of the Poisson bracket operator 
is given by 
\begin{eqnarray}
\left\{ \, \, \, , \, \, \right\}_{(\widehat{A},\widehat{B})} 
\xrightarrow[\hbar \rightarrow 0]{}   
\frac{\partial }{\partial A} \frac{\partial }{\partial B} - \frac{\partial }{\partial B} \frac{\partial }{\partial A} 
\, ,
\end{eqnarray}
where $A$ and $B$ are the classical counterparts of $\widehat{A}$ and $\widehat{B}$, respectively. 
For the right-hand side to reproduce the Poisson bracket, 
the operators $\widehat{A}$ and $\widehat{B}$ are identified with 
the position and momentum operators, respectively.

When the commutation relation for the canonical operators is given by 
a non-vanishing c-number, 
\begin{eqnarray}
[\widehat{A},\widehat{B}] = c \, , \label{eqn:commu_const}
\end{eqnarray}
we can show that the Poisson bracket operator is represented by the commutator,  
\begin{eqnarray}
[f(\widehat{A},\widehat{B}) , g(\widehat{A},\widehat{B}) ] 
=
c \left\{ f (\widehat{A},\widehat{B}), g(\widehat{A},\widehat{B}) 
\right\}_{(\widehat{A},\widehat{B})} 
\, .\label{eqn:com_poi}
\end{eqnarray}

{\bf Proof} 

Let us expand $f(\widehat{A},\widehat{B})$ and $g(\widehat{A},\widehat{B})$ as 
\begin{eqnarray}
\begin{split}
f (\widehat{A},\widehat{B}) &= \sum_{n,m \ge 0} f_{nm} \widehat{A}^n \widehat{B}^m \, , \\
g (\widehat{A},\widehat{B}) &= \sum_{n,m \ge 0} g_{nm} \widehat{A}^n \widehat{B}^m \, ,
\end{split}
\end{eqnarray} 
where $n$ and $m$ are non-negative integers and $f_{nm}$ and $g_{nm}$ are expansion coefficients. 
The commutator of these operators is calculated by
\begin{eqnarray}
\lefteqn{[f (\widehat{A},\widehat{B})  , g (\widehat{A},\widehat{B}) ]} && \nonumber \\
&=& 
\sum_{a,b,c,d \ge 0} f_{ab} g_{cd}
\left(
\widehat{A}^a \widehat{B}^b \widehat{A}^c \widehat{B}^d 
- 
\widehat{A}^c \widehat{B}^d \widehat{A}^a \widehat{B}^b 
\right)
\nonumber \\
&=& 
\sum_{a,b,c,d\ge 0} f_{ab} \, g_{cd}\, \widehat{A}^a \left[
bc (\delta_{\widehat{B}} \widehat{A}) \left( \int^1_0 \ud \lambda \, (\widehat{B} - \lambda \delta_{\widehat{B}})^{b-1} \widehat{A}^{c-1}\right) 
+ \widehat{A}^c \widehat{B}^b 
\right]\widehat{B}^d \nonumber \\
&&-
\sum_{a,b,c,d\ge 0} f_{ab} \,  g_{cd}\, \widehat{A}^c \left[
ad (\delta_{\widehat{B}} \widehat{A}) \left( \int^1_0 \ud \lambda \, (\widehat{B} - \lambda \delta_{\widehat{B}})^{d-1} \widehat{A}^{a-1} \right)
+ \widehat{A}^a \widehat{B}^d 
\right]\widehat{B}^b \, .
\end{eqnarray}
Here we used the formula (\ref{eqn:mformula3}).
Applying the formula (\ref{eqn:mformula2}) to the last line, 
we find 
\begin{eqnarray}
\lefteqn{[f (\widehat{A},\widehat{B})  , g (\widehat{A},\widehat{B}) ]} && \nonumber \\
&=& 
\sum_{a,b,c,d\ge 0} f_{ab} \,
\widehat{A}^a 
b (\delta_{\widehat{B}} \widehat{A}) \left( \int^1_0 \ud \lambda \,(\widehat{B} - \lambda \delta_{\widehat{B}})^{b-1} \frac{\ud g( \widehat{A}, 
\widehat{B}) }{\ud \widehat{A}} \right) \nonumber \\
&&-
\sum_{a,b,c,d\ge 0} f_{ab} \,
 a (\delta_{\widehat{B}} \widehat{A}) \left( \int^1_0 \ud \lambda \,(\widehat{A} - \lambda \delta_{\widehat{A}})^{a-1} \frac{\ud g(\widehat{A}, \widehat{B})}{\ud \widehat{B}} \right) 
\widehat{B}^b \nonumber \\
&=& 
(\delta_{\widehat{B}} \widehat{A}) \frac{\ud f(\widehat{A}, \widehat{B})}{\ud \widehat{B}} \left\{ \frac{\ud g(\widehat{A}, 
\widehat{B}) }{\ud \widehat{A}}\right\}
- (\delta_{\widehat{B}} \widehat{A}) \frac{\ud f(\widehat{A}, \widehat{B})}{\ud \widehat{A}} \left\{ \frac{\ud g(\widehat{A}, 
\widehat{B}) }{\ud \widehat{B}}\right\} \nonumber \\
&=& 
- (\delta_{\widehat{B}} \widehat{A})
\left\{ f( \widehat{A}, \widehat{B}), g(\widehat{A}, 
\widehat{B}) 
\right\}_{(\widehat{A},\widehat{B})} \, .
\end{eqnarray}
In the present case, $\delta_{\widehat{B}} \widehat{A} = -c \neq 0$. Therefore Eq.\ (\ref{eqn:com_poi}) was derived.
$\square$

The above result can be generalized to many-body systems.
We consider $N$-pairs of canonical variables $(\widehat{A}_i,\widehat{B}_i)$ 
which satisfy the following commutation relations for $i,j=1,\cdots N$,
\begin{eqnarray}
\begin{split}
[\widehat{A}_i, \widehat{B}_j  ] &= c_i \, \delta_{ij} \, ,\\
{\protect [\widehat{A}_i,\widehat{A}_j]} &= 0 \, ,\\
{\protect [\widehat{B}_i,\widehat{B}_j]} &= 0 \, ,
\end{split}
\label{eqn:cond_eq_bra_com}
\end{eqnarray}
where $c_i$ are c-numbers. 
Let us consider two smooth functions $f(\{ \widehat{A},\widehat{B} \} ) $ and $g(\{ \widehat{A},\widehat{B} \} ) $ 
which can be expanded as
\begin{eqnarray}
f(\{ \widehat{A},\widehat{B} \} ) 
&=& \sum_{\alpha_1,\beta_1 \cdots \alpha_N,\beta_N \ge 0} 
f_{\alpha_1, \beta_1 \cdots \alpha_N,\beta_N} \widehat{A}_1^{\alpha_1}  \widehat{B}_1^{\beta_1} \cdots 
\widehat{A}_N^{\alpha_N} \widehat{B}_N^{\beta_N} \, , 
\label{eqn:gen_op_f} \\
g(\{ \widehat{A},\widehat{B} \} ) 
&=& \sum_{\alpha_1,\beta_1 \cdots \alpha_N,\beta_N \ge 0}
g_{\alpha_1, \beta_1 \cdots \alpha_N,\beta_N} \widehat{A}_1^{\alpha_1}  \widehat{B}_1^{\beta_1}  \cdots 
\widehat{A}_N^{\alpha_N}  \widehat{B}_N^{\beta_N}  \, , 
\end{eqnarray}
where 
$\alpha_1,\beta_1 \cdots \alpha_N,\beta_N$ are non-negative integers and 
$f_{\alpha_1, \beta_1 \cdots \alpha_N,\beta_N}$ and $g_{\alpha_1, \beta_1 \cdots \alpha_N,\beta_N}$ 
are expansion coefficients.
Then the commutator is expressed as 
\begin{eqnarray}
\left[
f(\{ \widehat{A},\widehat{B} \} ) 
\, ,\, 
g(\{ \widehat{A},\widehat{B} \} ) 
\right] 
= 
\sum_{i=1}^N c_i  \left\{
f(\{ \widehat{A},\widehat{B} \} ) 
\, ,\, 
g(\{ \widehat{A},\widehat{B} \} ) 
\right\}_{(\widehat{A}_i,\widehat{B}_i )} \, .
\label{eqn:manybody}
\end{eqnarray}

{\bf Proof}

It is sufficient to prove the following equation,  
\begin{eqnarray}
\lefteqn{\left[
\widehat{A}_1^{\alpha_1}  \widehat{B}_1^{\beta_1}  \cdots 
\widehat{A}_N^{\alpha_N}  \widehat{B}_N^{\beta_N}  
\, , \, 
\widehat{A}_1^{\gamma_1}  \widehat{B}_1^{\delta_1}  \cdots 
\widehat{A}_N^{\gamma_N} \widehat{B}_N^{\delta_N}  
\right]} && \nonumber \\
&=& 
\sum_{i=1}^N c_i  \left\{
\widehat{A}_1^{\alpha_1}  \widehat{B}_1^{\beta_1}  \cdots 
\widehat{A}_N^{\alpha_N}  \widehat{B}_N^{\beta_N}  
\, , \, 
\widehat{A}_1^{\gamma_1}  \widehat{B}_1^{\delta_1}  \cdots 
\widehat{A}_N^{\gamma_N}  \widehat{B}_N^{\delta_N}  
\right\}_{(\widehat{A}_i,\widehat{B}_i)} \, .
\label{eqn:abababab}
\end{eqnarray}
The case for $N=1$ is already shown in Eq.\ (\ref{eqn:com_poi}).
Suppose that the above equation is satisfied for $N=L$ ($L \ge 1$).
Then, for $N=L+1$, we find
\begin{eqnarray}
\lefteqn{[
\widehat{A}^{\alpha_1}_1 \widehat{B}^{\beta_1}_1\cdots \widehat{A}^{\alpha_{L+1}}_{L+1} \widehat{B}^{\beta_{L+1}}_{L+1}
\, , \, 
\widehat{A}^{\gamma_1}_1\widehat{B}^{\delta_1}_1 \cdots \widehat{A}^{\gamma_{L+1}}_{L+1}\widehat{B}^{\delta_{L+1}}_{L+1}
 ]} \nonumber \\
&=& F^{(L)}_{AB} G^{(L)}_{AB} [ \widehat{A}^{\alpha_{L+1}}_{L+1}\widehat{B}^{\beta_{L+1}}_{L+1}
\, , \,   \widehat{A}^{\gamma_{L+1}}_{L+1} \widehat{B}^{\delta_{L+1}}_{L+1} ]
+ [F^{(L)}_{AB} \, , \,   G^{(L)}_{AB}]\widehat{A}^{\gamma_{L+1}}_{L+1} \widehat{B}^{\delta_{L+1}}_{L+1} \widehat{A}^{\alpha_{L+1}}_{L+1}\widehat{B}^{\beta_{L+1}}_{L+1} \nonumber \\
&=& 
c_{L+1} F^{(L)}_{AB} G^{(L)}_{AB} 
\left\{
 \widehat{A}^{\alpha_{L+1}}_{L+1}  \widehat{B}^{\beta_{L+1}}_{L+1} 
\, , \, 
\widehat{A}^{\gamma_{L+1}}_{L+1} \widehat{B}^{\delta_{L+1}}_{L+1}
\right\}
_{(\widehat{A}_{L+1},\widehat{B}_{L+1})} \nonumber \\
&& + 
\sum_{i=1}^L c_i \left\{
F^{(L)}_{AB} \, , \,   G^{(L)}_{AB}\right\}_{(\widehat{A}_i,\widehat{B}_i)} \widehat{A}^{\gamma_{L+1}}_{L+1}\widehat{B}^{\delta_{L+1}}_{L+1} 
\widehat{A}^{\alpha_{L+1}}_{L+1}  \widehat{B}^{\beta_{L+1}}_{L+1}
\nonumber \\
&=&  \sum_{i=1}^{L+1} c_i  \left\{
\widehat{A}^{\alpha_1}_1\cdots \widehat{A}^{\alpha_L}_L\widehat{B}^{\beta_1}_1\cdots \widehat{B}^{\beta_L}_L
\, , \, 
\widehat{A}^{\gamma_1}_1\cdots \widehat{A}^{\gamma_{L+1}}_{L+1}\widehat{B}^{\delta_1}_1\cdots \widehat{B}^{\delta_{L+1}}_{L+1}
\right\}_{(\widehat{A}_i,\widehat{B}_i)} \, ,
\end{eqnarray}
where we introduced 
\begin{eqnarray}
\begin{split}
F^{(L)}_{AB} &= \widehat{A}^{\alpha_1}_1 \widehat{B}^{\delta_1}_1\cdots \widehat{A}^{\alpha_{L}}_{L} \widehat{B}^{\delta_{L}}_{L}\, , \\
G^{(L)}_{AB} &= \widehat{A}^{\gamma_1}_1 \widehat{B}^{\delta_1}_1\cdots \widehat{A}^{\gamma_{L}}_{L}  \widehat{B}^{\delta_{L}}_{L} \, .
\end{split}
\end{eqnarray}
From the first to the second equality, we used Eq.\ (\ref{eqn:abababab}) by mathematical induction.
The last equality is the right-hands side of Eq.\ (\ref{eqn:abababab}) for $N= L+1$ and thus 
Eq.\ (\ref{eqn:abababab}) is hold for arbitrary integer $N \ge 1$.
From this, it is easy to show the formula (\ref{eqn:manybody}).
$\square$

From the definition of the operator derivative, we can show that the Poisson bracket operator satisfies the following properties,
\begin{eqnarray}
\lefteqn{\left\{
a\, f(\{ \widehat{A},\widehat{B} \} ) + b\, g(\{ \widehat{A},\widehat{B} \} ) 
\, ,\, 
h(\{ \widehat{A},\widehat{B} \} ) 
\right\}_{(\{ \widehat{A},\widehat{B}\} )}} && \nonumber \\
&=& 
 a \left\{
f(\{ \widehat{A},\widehat{B} \} ) 
\, ,\, 
h(\{ \widehat{A},\widehat{B} \} ) 
\right\}_{(\{\widehat{A},\widehat{B}\} )}
+
 b \left\{
g(\{ \widehat{A},\widehat{B} \} ) 
\, ,\, 
h(\{ \widehat{A},\widehat{B} \} ) 
\right\}_{(\{\widehat{A},\widehat{B}\} )} \, ,\\
\lefteqn{
 \left\{
f(\{ \widehat{A},\widehat{B} \} ) g (\{ \widehat{A},\widehat{B} \} ) 
\, ,\, 
h(\{ \widehat{A},\widehat{B} \} ) 
\right\}_{(\{\widehat{A},\widehat{B}\} )}} && \nonumber \\
&=& 
 \left\{
f(\{ \widehat{A},\widehat{B} \} )  
\, ,\, 
h(\{ \widehat{A},\widehat{B} \} ) 
\right\}_{ (\{ \widehat{A},\widehat{B}\} )}
g (\{ \widehat{A},\widehat{B} \} )
+
f (\{ \widehat{A},\widehat{B} \} )
 \left\{
g(\{ \widehat{A},\widehat{B} \} )  
\, ,\, 
h(\{ \widehat{A},\widehat{B} \} ) 
\right\}_{(\{ \widehat{A},\widehat{B} \} )} \, , \nonumber \\
\end{eqnarray}
where $a$ and $b$ are constants and $h(\{ \widehat{A},\widehat{B} \} ) $ is another smooth function like $f(\{ \widehat{A},\widehat{B} \} ) $ and $g(\{ \widehat{A},\widehat{B} \} )$.
To simplify the equations, we introduced the notation,
\begin{eqnarray}
\left\{
f(\{ \widehat{A},\widehat{B} \} )  
\, ,\, 
g(\{ \widehat{A},\widehat{B} \} ) 
\right\}_{(\{ \widehat{A},\widehat{B}\} )}
\equiv
\sum_{i=1}^N \left\{
f(\{ \widehat{A},\widehat{B} \} )  
\, ,\, 
g(\{ \widehat{A},\widehat{B} \} ) 
\right\}_{(\widehat{A}_i,\widehat{B}_i )} \, .
\end{eqnarray}

Other two properties for the Poisson bracket operator are however shown 
using the condition (\ref{eqn:cond_eq_bra_com}) and $c_i = \ii \hbar$.
In this case, we find 
\begin{eqnarray}
-\frac{\ii}{\hbar} \left[
f(\{ \widehat{A},\widehat{B} \} ) 
\, ,\, 
g(\{ \widehat{A},\widehat{B} \} ) 
\right]  = 
\left\{
f(\{ \widehat{A},\widehat{B} \} ) 
\, ,\, 
g(\{ \widehat{A},\widehat{B} \} ) 
\right\}_{(\{ \widehat{A},\widehat{B}\} )} \, .
\end{eqnarray}
Therefore it is easy to confirm that the Poisson bracket operator satisfies
\begin{eqnarray}
 \left\{
f(\{ \widehat{A},\widehat{B} \} ) 
\, ,\, 
g(\{ \widehat{A},\widehat{B} \} ) 
\right\}_{(\{ \widehat{A},\widehat{B}\} )}
= 
- 
 \left\{
g(\{ \widehat{A},\widehat{B} \} ) 
\, ,\, 
f(\{ \widehat{A},\widehat{B} \} ) 
\right\}_{(\{ \widehat{A},\widehat{B}\} )} \, ,
\label{eqn:interchange}
\end{eqnarray}
and the Jacobi identity 
\begin{eqnarray}
&& \left\{
f(\{ \widehat{A},\widehat{B} \} ) 
\, ,\, 
\left\{
g(\{ \widehat{A},\widehat{B} \} ) 
\, \, 
h(\{ \widehat{A},\widehat{B} \} ) 
\right\}_{(\{ \widehat{A},\widehat{B}\} )}
\right\}_{(\{ \widehat{A},\widehat{B}\} )} \nonumber \\
&&+
 \left\{
g(\{ \widehat{A},\widehat{B} \} ) 
\, ,\, 
\left\{
h(\{ \widehat{A},\widehat{B} \} ) 
\, \, 
f(\{ \widehat{A},\widehat{B} \} ) 
\right\}_{(\{ \widehat{A},\widehat{B}\} )}
\right\}_{(\{ \widehat{A},\widehat{B}\} )} \nonumber \\
&&+
 \left\{
h(\{ \widehat{A},\widehat{B} \} ) 
\, ,\, 
\left\{
f(\{ \widehat{A},\widehat{B} \} ) 
\, \, 
g(\{ \widehat{A},\widehat{B} \} ) 
\right\}_{(\{ \widehat{A},\widehat{B}\} )}
\right\}_{(\{ \widehat{A},\widehat{B}\} )}
=0 \, .
\end{eqnarray}
We further confirmed that the above two equations are satisfied in several examples where the condition (\ref{eqn:cond_eq_bra_com}) is not applicable, but 
the general proofs are not known.
See also the discussion in Appendix\ \ref{app:conserve}.

\section{Quantum canonical equation}\label{sec:qc_eq}

Let us consider the system which is described by $N$-pairs of canonical variables $(\widehat{A}_i (t), \widehat{B}_i (t))$, 
$(i = 1,\cdots N)$.
In quantum mechanics, the time evolutions of operators are described by the Heisenberg equation,  
\begin{eqnarray}
\frac{\ud}{\ud t} f(\{ \widehat{A}(t),\widehat{B}(t) \}) 
= 
-\frac{\ii}{\hbar} [ f(\{ \widehat{A}(t),\widehat{B}(t) \}) , \widehat{H} ]
\, ,
\end{eqnarray}
where $\widehat{H}$ is the Hamiltonian operator.
The commutation relations of the canonical variables are characterized by the same constant $\ii \hbar$, 
\begin{eqnarray}
\begin{split}
[\widehat{A}_i (t),\widehat{B}_j (t)] &= \ii \hbar \, \delta_{ij} \, , \\
{\protect [\widehat{A}_i (t),\widehat{A}_j (t)]} &= 0 \, , \\
{\protect [\widehat{B}_i (t),\widehat{B}_j (t)]} &= 0 \, .
\end{split}
\label{eqn:standard_commu}
\end{eqnarray}  
Using the property (\ref{eqn:manybody}) with $c_i = \ii \hbar$, 
we can reexpress the right-hand side of the Heisenberg equation in terms of the Poisson bracket operator, 
\begin{eqnarray}
\frac{\ud}{\ud t} f(\{ \widehat{A}(t),\widehat{B}(t) \}) 
=
\left\{ f(\{ \widehat{A}(t),\widehat{B}(t) \}), \widehat{H} \right\}_{(\{\widehat{A} (t), \widehat{B} (t)\})} \, .
\label{eqn:qce}
\end{eqnarray} 
We call this the quantum canonical equation. 
As will be seen later, the quantum canonical equation is not necessarily equivalent to the Heisenberg equation 
in non-standard applications.

The quantum canonical equation is consistent with the mathematical property of the operator derivative. 
From Eqs. (\ref{eqn:def_tderi}) and (\ref{eqn:def_par_deri}), 
the time derivative of $ f(\{ \widehat{A}(t),\widehat{B}(t) \}) $ is given by  
\begin{eqnarray}
\frac{\ud}{\ud t} f(\{ \widehat{A}(t),\widehat{B}(t) \}) 
&=& 
\sum_{i=1}^N \frac{\partial f(\{ \widehat{A}(t),\widehat{B}(t) \}) }{\partial \widehat{A}_i (t)} \left\{ \frac{\ud \widehat{A}_i (t)}{\ud t}  \right\}
+
\sum_{i=1}^N \frac{\partial f(\{\widehat{A}(t),\widehat{B}(t) \}) }{\partial \widehat{B}_i (t)} \left\{ \frac{\ud \widehat{B}_i (t)}{\ud t}  \right\}
\, . \label{eqn:1}
\nonumber \\
\end{eqnarray}
Therefore one can see that the above equation reproduces the quantum canonical equation when 
$\widehat{A}_i (t)$ and $\widehat{B}_i (t)$ satisfies the Heisenberg equations,  
\begin{eqnarray}
\begin{split}
\frac{\ud \widehat{A}_i (t)}{\ud t} 
&
= 
\left\{ \widehat{A}(t), \widehat{H} \right\}_{(\{\widehat{A} (t), \widehat{B} (t)\})}
= \frac{\partial \widehat{H} }{\partial \widehat{B}_i (t)} 
\, ,\\
\frac{\ud \widehat{B}_i (t)}{\ud t} 
&
= 
\left\{ \widehat{B}(t), \widehat{H} \right\}_{(\{\widehat{A} (t), \widehat{B} (t)\})}
= - \frac{\partial \widehat{H} }{\partial \widehat{A}_i (t)} 
\, .
\end{split}
\label{eqn:ehrenfest}
\end{eqnarray}

The correspondence between classical and quantum behaviors is clear in the quantum canonical equation. 
The Poisson bracket operator is defined independently of the property of the commutation relation $[\widehat{A}_i (t),\widehat{B}_j (t)]$ and thus 
the quantum canonical equation is applicable to the commutative case, $[\widehat{A}_i (t),\widehat{B}_j (t)]=0$. 
Because the Poisson bracket operator behaves as the Poisson bracket for c-number variables, 
the quantum canonical equation reproduces the classical canonical equation in the application to c-number canonical variables. 
In other words, the quantum canonical equation enables us to describe classical and quantum behaviors in a unified way.

In the operator derivative (\ref{eqn:def_op_der}), 
the effect of non-commutativity is represented through $\delta_{\widehat{A}}$.
To see the quantum effect in the quantum canonical equation clearly, we represent it 
in the series expansion of $\delta_{\widehat{A}}$.
As an example, we consider a single-particle system described by the Hamiltonian operator 
\begin{eqnarray}
\widehat{H} = \frac{\widehat{p}^2_t}{2\um} + V(\widehat{x}_t) \, ,
\end{eqnarray}
where $V(x)$ is the potential energy 
and the canonical operators satisfy the standard canonical commutation relation, 
\begin{eqnarray}
[\widehat{x}_t,\widehat{p}_t] = \ii \hbar \, .
\end{eqnarray}
The quantum canonical equation for $f(\widehat{x}_t,\widehat{p}_t)$ is given by 
\begin{eqnarray}
\frac{\ud}{\ud t} f(\widehat{x}_t,\widehat{p}_t) = \left\{ f(\widehat{x}_t,\widehat{p}_t), \widehat{H}\right\}_{(\widehat{x}_t,\widehat{p}_t)} 
= - \left\{ \widehat{H}, f(\widehat{x}_t,\widehat{p}_t)\right\}_{(\widehat{x}_t,\widehat{p}_t)} \, .
\end{eqnarray}
Here we used Eq.\ (\ref{eqn:interchange}).
Then the right-hand side of the quantum canonical equation can be expanded as
\begin{eqnarray}
\frac{\ud f(\widehat{x}_t,\widehat{p}_t)}{\ud t} 
&=&
\frac{\widehat{p}_t}{\um} \frac{\partial f(\widehat{x}_t,\widehat{p}_t)}{\partial \widehat{x}_t} - V^{(1)}(\widehat{x}_t) \frac{\partial f(\widehat{x}_t,\widehat{p}_t)}{\partial \widehat{p}_t} \nonumber \\
&& - \frac{1}{2\um} \delta_{\widehat{p}_t} \frac{\partial f(\widehat{x}_t,\widehat{p}_t)}{\partial \widehat{x}_t} 
- \sum_{m=1}^\infty \frac{1}{(m+1)!} V^{(m+1)}(\widehat{x}_t) (-\delta_{\widehat{x}_t})^{m} \frac{\partial f(\widehat{x}_t,\widehat{p}_t)}{\partial \widehat{p}_t}
\, . \label{eqn:deltaexp}
\end{eqnarray}
In the classical limit, only the first two terms survive on the right-hand side 
and then it is easy to see that the classical canonical equation is reproduced.

When the standard canonical commutation relation is satisfied, there exist the following relations,
\begin{eqnarray}
\begin{split}
\delta_{\widehat{x}_t} f(\widehat{x}_t,\widehat{p}_t) &= \ii \hbar \frac{\partial f(\widehat{x}_t,\widehat{p}_t)}{\partial \widehat{p}_t} \, ,\\
\delta_{\widehat{p}_t} f(\widehat{x}_t,\widehat{p}_t) &= -\ii \hbar \frac{\partial f(\widehat{x}_t,\widehat{p}_t)}{\partial \widehat{x}_t} \, .
\end{split}\label{eqn:delta_deri}
\end{eqnarray}
Using these, Eq.\ (\ref{eqn:deltaexp}) can be reexpressed in various different forms. 
One expression is
\begin{eqnarray}
\ii \hbar \frac{\ud f(\widehat{x}_t,\widehat{p}_t)}{\ud t} 
&=&
 \left[ \frac{(\widehat{p}_t - \delta_{\widehat{p}_t})^2}{2\um} + V(\widehat{x}_t-\delta_{\widehat{x}_t}) 
- \frac{\widehat{p}_t^2}{2\um}
- V(\widehat{x}_t) \right] f(\widehat{x}_t,\widehat{p}_t)
\, .
\end{eqnarray}
See also Eqs.\ (3.12) and (3.13) in Ref.\ \cite{qa6}.
This representation is compact but not suitable to see the classical limit.

The other expression is given by
\begin{eqnarray}
\frac{\ud f(\widehat{x}_t,\widehat{p}_t)}{\ud t} 
&=&
\frac{\widehat{p}_t}{\um} \frac{\partial f(\widehat{x}_t,\widehat{p}_t)}{\partial \widehat{x}_t}  
+ \frac{\ii \hbar}{2\um} \frac{\partial^2 f(\widehat{x}_t,\widehat{p}_t)}{\partial \widehat{x}^2_t} 
\nonumber \\
&&
- \sum_{m=0}^\infty \frac{1}{(m+1)!} V^{(m+1)}(\widehat{x}_t) 
\left( - \ii \hbar \right)^m
 \frac{\partial^{m+1} f(\widehat{x}_t,\widehat{p}_t)}{\partial \widehat{p}^{m+1}_t} \, ,
\label{eqn:qce_expa}
\end{eqnarray}
where the higher-order operator derivatives are operated to c-numbers and thus 
\begin{eqnarray}
\begin{split}
\frac{\partial^n f(\widehat{x}_t,\widehat{p}_t)}{\partial \widehat{x}^n_t} 
&=
\frac{\partial^n f(\widehat{x}_t,\widehat{p}_t)}{\partial \widehat{x}^n_t} \{ 1\} 
= \left. \frac{\partial^n f(x,\widehat{p}_t)}{\partial x^n} \right|_{x=\widehat{x}_t} \, ,\\
\frac{\partial^n f(\widehat{x}_t,\widehat{p}_t)}{\partial \widehat{p}^n_t} 
&=
\frac{\partial^n f(\widehat{x}_t,\widehat{p}_t)}{\partial \widehat{p}^n_t} \{ 1\} 
= \left. \frac{\partial^n f(\widehat{x}_t,p)}{\partial p^n} \right|_{p=\widehat{p}_t} \, .
\end{split}
\end{eqnarray}
For the definition of the higher-order operator derivative as a hyper-operator, see Eq. (2.16) in Ref.\ \cite{qa7}.
In this expression, $\delta_{\widehat{x}_t}$ and $\delta_{\widehat{p}_t}$ are replaced with the operator derivatives. We see that the order of the quantum correction and that of the operator derivative are correlated.

This expansion reminds us of the evolution equation of the Wigner function \cite{wigner}.
To see this, we choose 
\begin{eqnarray}
f(\widehat{x}_t,\widehat{p}_t) = \delta (x-\widehat{x}_t+ \delta_{\widehat{x}_t}/2) \delta (p-\widehat{p}_t) \, ,
\label{eqn:wigner_choice}
\end{eqnarray}
where
\begin{eqnarray}
\delta(z) = \frac{1}{2\pi} \int \ud k\, e^{\ii k z} \, .
\end{eqnarray}
Then the Wigner function is defined by 
\begin{eqnarray}
W(x,p,t) 
&=& \langle \psi |  \delta (x-\widehat{x}_t+ \delta_{\widehat{x}_t}/2) \delta (p-\widehat{p}_t) | \psi \rangle 
\nonumber \\
&=&
\frac{1}{2\pi \hbar} \int \ud q\, \psi^* (x+q/2,t) \psi(x-q/2,t) e^{\ii pq/\hbar} 
\, , 
\label{eqn:wfunction}
\end{eqnarray}
where $| \psi \rangle$ is an initial wave function and $\psi (x,t) = \langle x| e^{-\ii \widehat{H} t/\hbar} |\psi \rangle$.
Using these definitions in Eq.\ (\ref{eqn:qce_expa}), the well-known evolution equation of the Wigner function 
is reproduced, 
\begin{eqnarray}
\partial_t W(x,p,t) 
&=&
-\frac{p}{m} \partial_x W(x,p,t) + V^{(1)} (x) \partial_p W(x,p,t) \nonumber \\
&& + \sum_{l=1}^\infty \frac{V^{(2l+1)}(x)}{(2l+1)!} \left( - \frac{\hbar^2}{4}  \right)^{l} \partial^{2l+1}_p W(x,p,t) 
\, . 
\label{eqn:eq_wigner}
\end{eqnarray}
See also the discussions in Refs.\ \cite{qse_koide,qse_koide2}.
Differently from Eq.\ (\ref{eqn:qce_expa}), Eq.\ (\ref{eqn:eq_wigner})  is given by c-numbers and  
only the odd-order terms of the momentum derivative appear on the right-hand side 
because of the property of Eq.\ (\ref{eqn:wigner_choice}). 
The constant factor $(1/4)^l$ in Eq.\ (\ref{eqn:eq_wigner}) is reproduced  
when the operator derivatives are replaced with the c-number derivatives in Eq.\ (\ref{eqn:qce_expa}).
Therefore Eq.\ (\ref{eqn:qce_expa}) can be regarded as the operator-derivative representation of the Moyal bracket 
in the Heisenberg equation when it is applied to the Wigner function.

In our approach, the classical derivatives with respect to position and momentum 
are replaced with the corresponding 
operator derivatives in the quantization of the Poisson bracket.
This picture may be utilized to extend the idea of quantization.
The diffusion equation describes a typical dissipative phenomenon in classical systems,
\begin{eqnarray}
\frac{\partial}{\partial t} \rho = D \frac{\partial^2}{\partial x^2} \rho \, ,
\end{eqnarray}
where $\rho$ is a conserved density normalized by one and $D$ is a diffusion constant. 
Suppose that this equation can be quantized by replacing $\rho$ and $\partial/\partial x$ 
with the density matrix $\widehat{\rho}$ and the operator derivative $\partial/\partial {\widehat{x}}$, respectively. 
The derived equation is given by 
\begin{eqnarray}
\frac{\partial}{\partial t}  \widehat{\rho} = D \frac{\partial^2}{\partial \widehat{x}^2} \widehat{\rho} 
= -\frac{D}{\hbar^2} \delta^2_{\widehat{p}} \widehat{\rho} \, .
\label{eqn:lindblad}
\end{eqnarray}
In the second equality, we used Eq.\ (\ref{eqn:delta_deri}).
This equation can be reexpressed as 
\begin{eqnarray}
\partial_t \widehat{\rho} = -\frac{1}{2} (\widehat{L}^2 \widehat{\rho} + \widehat{\rho} \widehat{L}^2) 
+ \widehat{L} \widehat{\rho} \widehat{L} \, ,
\end{eqnarray}
where 
\begin{eqnarray}
\widehat{L} = \sqrt{\frac{2D}{\hbar}} \widehat{p} \, .
\end{eqnarray}
That is, Eq.\ (\ref{eqn:lindblad}) reproduces the Lindblad equation \cite{lindblad,gorini}. 

Note that higher-order operator-derivative terms can be induced in the quantization of the Poisson bracket 
as seen from Eq.\ (\ref{eqn:qce_expa}).
This is however not necessarily applicable to the diffusion equation because 
the canonical equations are not established in dissipative systems \cite{dekker,bagarello,um}. 
See also Refs.\ \cite{koide17,koide18,zambrini1,zambrini2} to find other interesting relations between the classical diffusion and quantum mechanics.

\section{Application to non-standard system} \label{sec:app}

As was shown in the previous section, the Poisson bracket operator is equivalent to the commutator when canonical variables satisfy 
the standard canonical commutation relations (\ref{eqn:standard_commu}).
The quantum canonical equation is definable independently of the behaviors of commutation relations 
and thus is applicable to a system where the Heisenberg equation is not defined. 
As an example, we consider a system where a c-number and a q-number particles coexist.

\subsection{model and application of quantum canonical equation }

As an example of the non-standard application, we consider a system where two pairs of canonical variables 
satisfy different commutation relations ($i,j=1,2$), 
\begin{eqnarray}
[\widehat{A}_i (t), \widehat{B}_j (t)] &= c_i (t) \, \delta_{ij} \, , 
\end{eqnarray}
where $c_1 (t)$ and $c_2 (t)$ are c-numbers and $c_1 (t) \neq c_2 (t)$.
Suppose that the system 1 described by $(\widehat{A}_1 (t), \widehat{B}_1(t))$ is separated from the system 2 by $(\widehat{A}_2 (t), \widehat{B}_2(t))$, and these systems do not interact each other.
The Heisenberg equations for the system 1 and for the system 2 will be, respectively, characterized by $c_1(t)$ and $c_2 (t)$, 
\begin{eqnarray}
\frac{\ud }{\ud t} f (\widehat{A}_1(t) ,\widehat{B}_1(t) ) &= \frac{1}{c_1(t)} [ f (\widehat{A}_1(t) ,\widehat{B}_1 (t) ), \widehat{H}_1 ] \, ,  \label{eqn:sys_h1} \\
\frac{\ud }{\ud t} f (\widehat{A}_2(t) ,\widehat{B}_2(t)) &= \frac{1}{c_2(t)} [ f (\widehat{A}_2(t) ,\widehat{B}_2(t)), \widehat{H}_2 ] \, . \label{eqn:sys_h2}
\end{eqnarray}
Here $\widehat{H}_1$ and $\widehat{H}_2$ are the Hamiltonian operators for each system.
When an interaction between the two systems is introduced, 
we have to generalize the Heisenberg equations so that 
the generalized equation reproduces Eqs.\ (\ref{eqn:sys_h1}) and (\ref{eqn:sys_h2})  in the vanishing limit of the interaction. 
Such a generalization is not trivial.

By contrast, the quantum canonical equation is applicable to such a system systematically. 
As an extreme case of the above system, we consider a toy model where two particles coexist: 
one is described by a c-number and the other by a q-number when there is no interaction between the particles.
The canonical variables for the former particle are denoted by $(x(t),p(t))$ and those for the latter by $(\widehat{X}(t),\widehat{P}(t))$.
The commutation relations are thus given by 
\begin{eqnarray}
\begin{split}
[x(t), p(t)] &= 0 \, ,\\
{\protect [\widehat{X}(t), \widehat{P}(t)]} &= \ii \hbar \, ,
\end{split}
\label{eqn:commu_noint}
\end{eqnarray}
respectively.
For the sake of simplicity, we assume free particles. 
Then the c-number canonical variables satisfy the classical canonical equations, 
\begin{eqnarray}
\begin{split}
\frac{\ud}{\ud t} x(t) &= \{ x(t) , H_{c}  \}_{PB}  = \frac{p(t)}\um{}\, , \\
\frac{\ud}{\ud t} p(t) &= \{ p(t) , H_{c}  \}_{PB} = 0 \, ,
\end{split}
\label{eqn:noint1}
\end{eqnarray}
where the c-number Hamiltonian with mass $\um$ is defined by 
\begin{eqnarray}
H_{c} = \frac{p^2(t)}{2\um} \, .
\end{eqnarray}
The q-number canonical variables satisfy the Heisenberg equations,
\begin{eqnarray}
\begin{split}
\frac{\ud}{\ud t} \widehat{X}(t) 
&= 
-\frac{\ii}{\hbar} [\widehat{X}(t)  , \widehat{H}_{q} ] = \frac{\widehat{P}(t)}{\uM}
\, , \\
\frac{\ud}{\ud t} \widehat{P}(t) 
&= 
-\frac{\ii}{\hbar} [\widehat{P}(t)  , \widehat{H}_{q} ] = 0\, ,
\end{split}
\label{eqn:noint2}
\end{eqnarray}
where the q-number Hamiltonian with mass $\uM$ is 
\begin{eqnarray}
\widehat{H}_{q}  = \frac{\widehat{P}^2(t)}{2\uM} \, .
\end{eqnarray}

The above two equations are described by the common quantum canonical equation,
\begin{eqnarray}
\lefteqn{\frac{\ud}{\ud t} f (x(t), p(t),\widehat{X}(t), \widehat{P}(t)) } && \nonumber \\
&=& \{ f(x(t), p(t),\widehat{X}(t), \widehat{P}(t)) , \widehat{H} \}_{(x(t),p(t))}
 + 
\{ f(x(t), p(t),\widehat{X}(t), \widehat{P}(t)) , \widehat{H} \}_{(\widehat{X}(t),\widehat{P}(t))} \nonumber \\
&\equiv&
\{ f(x(t), p(t),\widehat{X}(t), \widehat{P}(t)) , \widehat{H} \}_{(x(t),p(t);\widehat{X}(t),\widehat{P}(t))}
\, ,
\label{eqn:qceq}
\end{eqnarray}
where $\widehat{H}$ is the total Hamiltonian defined by $\widehat{H} = H_{c} + \widehat{H}_{q}$, and 
$f(x(t), p(t),\widehat{X}(t), \widehat{P}(t))$ is a smooth function of the canonical variables which can be expanded 
as Eq.\ (\ref{eqn:gen_op_f}).
As pointed out earlier, the Poisson bracket operator behaves as the Poisson bracket for c-number variables.
In the following calculation,  we suppose that the quantum canonical equation is applicable 
to any Hamiltonian operator.

Let us consider the interaction Hamiltonian defined by 
\begin{eqnarray}
\widehat{H}_{I} = \frac{\alpha}{2} (x(t) - \widehat{X}(t))(x(t) - \widehat{X}(t))  \, , 
\end{eqnarray}
where $\alpha$ is a coupling constant. 
It should be noted that the canonical variables $(x(t),p(t))$ becomes non-commutative and behave as 
operators by the influence of this interaction.
Therefore the commutation relations (\ref{eqn:commu_noint}) are modified.
The modifications are obtained only after solving the quantum canonical equations 
as shown later.

Using the total Hamiltonian defined by   
\begin{eqnarray}
\widehat{H} = H_{c} + \widehat{H}_{q} + \widehat{H} _I \, , \label{eqn:totalh}
\end{eqnarray}
the quantum canonical equations are given by
\begin{eqnarray}
\begin{split}
\frac{\ud x(t)}{\ud t } &= \{ x(t) , \widehat{H} \}_{(x(t),p(t);\widehat{X}(t),\widehat{P}(t))} 
= \frac{p(t)}{\um}\, , \\
\frac{\ud p(t)}{\ud t } &= \{ p(t) , \widehat{H} \}_{(x(t),p(t);\widehat{X}(t),\widehat{P}(t))} 
= 
-\alpha (x(t) - \widehat{X}(t)) \, , \\
\frac{\ud \widehat{X}(t)}{\ud t } &= \{ \widehat{X}(t) , \widehat{H} \}_{(x(t),p(t);\widehat{X}(t),\widehat{P}(t))} 
= \frac{\widehat{P}(t)}{\uM} \, , \\
\frac{\ud \widehat{P}(t)}{\ud t } &= \{ \widehat{P}(t) , \widehat{H} \}_{(x(t),p(t);\widehat{X}(t),\widehat{P}(t))} 
 = \alpha (x(t) - \widehat{X}(t)) \, .
\end{split}
\label{eqn:model_eom}
\end{eqnarray}

To solve these equations, 
suppose that the two particles start to interact each other at an initial time $t=0$.
Then the initial canonical variables  
($x (0) = x_0$, $p (0) = p_0$) and ($\widehat{X}(0) = \widehat{X}_0$, $\widehat{P}(0) = \widehat{P}_0$) 
satisfy the following standard commutation relations, 
\begin{eqnarray}
\begin{split}
[x_0, p_0] &= 0 \, , \, \, \,  \, \, \, [\widehat{X}_0, \widehat{P}_0] = \ii \hbar \, ,\\
{\protect [x_0, \widehat{X}_0]} &= 0 \, , \, \, \,  \, \, \, [p_0, \widehat{P}_0] = 0 \, .
\end{split}
\label{eqn:ini_commu}
\end{eqnarray}
The canonical operators $\widehat{X}_0$ and $\widehat{P}_0$ operate to an initial 
wave function $|\Psi \rangle$ normalized by
\begin{eqnarray}
\langle \Psi | \Psi \rangle = 1 \, .
\end{eqnarray}

To solve the differential equations, we further introduce new canonical variables associated with the center of mass and relative motions, 
\begin{eqnarray}
( \widehat{X}_{C} (t) ,  \widehat{P}_{C}(t) ) &=& \left( \frac{\um x (t)+ \uM \widehat{X}(t)}{\um + \uM},  p (t) + \widehat{P}(t) \right) \, ,\\
( \widehat{q}(t), \widehat{p}_q (t) ) &=& 
\left(  x (t) - \widehat{X}(t) ,  \frac{\uM p (t)- \um\widehat{P}(t) }{\um+\uM} \right) \, .
\end{eqnarray}

The quantum canonical equations are simplified for these new canonical variables.
Solving the equations, the motions for the center of mass coordinates are described by  
\begin{eqnarray}
\begin{split}
\widehat{X}_C (t) &= \frac{\um x_0 + \uM \widehat{X}_0}{\um + \uM} + \frac{1}{\um + \uM} (p_0 + \widehat{P}_0) t \, , \\
\widehat{P}_C (t) &= p_0+ \widehat{P}_0 \, ,
\end{split}
\label{eqn:eq_com}
\end{eqnarray}
and those for the relative coordinates are 
\begin{eqnarray}
\begin{split}
\widehat{q}(t)
&= 
(x_0 - \widehat{X}_0) \cos \left( \sqrt{\frac{\alpha}{\mu} }t \right) 
+ \sqrt{\frac{\mu}{\alpha}} \left( \frac{p_0}{\um} - \frac{\widehat{P}_0}{\uM} \right) \sin \left( \sqrt{\frac{\alpha}{\mu} }t \right) \, ,\\
\widehat{p}_q (t)
&=
-\sqrt{\alpha \mu} (x_0 - \widehat{X}_0) \sin \left( \sqrt{\frac{\alpha}{\mu} }t \right) 
+ \left( \frac{\uM p_0 - \um \widehat{P}_0}{\um+\uM} \right) \cos \left( \sqrt{\frac{\alpha}{\mu} }t \right) \, .
\end{split}
\label{eqn:eq_qpq}
\end{eqnarray}
Here the reduced mass is defined  by 
\begin{eqnarray}
\mu = \frac{\um \uM}{\um +\uM} \, .
\end{eqnarray}

The Ehrenfest theorem and the non-interacting limit of this model are discussed in Appendix \ref{app:ehren}.
The momentum conservation is easily seen from Eq.\ (\ref{eqn:eq_com}). 
The total energy of the system is defined by the expectation value of the total Hamiltonian operator (\ref{eqn:totalh}) and this quantity is conserved, as is shown in Appendix \ref{app:conserve}.

\subsection{commutation relations} \label{sec:commu}

At the initial time $t=0$, the c-number and q-number particles behave as independent particles, 
satisfying the commutation relations (\ref{eqn:ini_commu}).
Because of the interaction, however, these commutation relations are modified.

The modified commutation relations are obtained from the solutions of the quantum canonical equations, (\ref{eqn:eq_com}) 
and (\ref{eqn:eq_qpq}).
The commutation relations for the pairs of the canonical variables are given by
\begin{eqnarray}
[\widehat{X}_G (t), \widehat{P}_G (t)]
&=& 
\frac{\uM}{\um + \uM} \ii \hbar \, , \\
{\protect [\widehat{q}(t), \widehat{p}_q(t)] }
&=&
\frac{\um}{\um+\uM} \ii \hbar \, .
\end{eqnarray}
The right-hand sides are characterized by different constants.

The commutation relations for the positions and for the momenta are 
\begin{eqnarray}
[\widehat{q}(t), \widehat{X}_G(t)]
&=& 
\frac{-\ii \hbar }{\um + \uM} 
\left[ 
t \cos \left( \sqrt{\frac{\alpha}{\mu}}t \right) 
-  \sqrt{\frac{\mu}{\alpha}}  \sin \left( \sqrt{\frac{\alpha}{\mu}}t \right) 
\right]
\, , \label{eqn:commu_qxg} \\
{\protect [ \widehat{p}_q (t) , \widehat{P}_G(t) ] }
&=& \ii \hbar  \sqrt{\alpha \mu} \sin \left( \sqrt{\frac{\alpha}{\mu} }t \right) \, ,
\end{eqnarray}
respectively.
One can easily see that the condition (\ref{eqn:cond_eq_bra_com}) is not satisfied in this toy model. 
Thus the Poisson bracket operator cannot be represented by the commutator and 
the quantum canonical equation does not coincide with the Heisenberg equation.
This means that the time evolution of operators cannot be represented by using the unitary operator.
Therefore this toy model is described only in the Heisenberg picture and the corresponding Schr\"{o}dinger picture 
is not defined.

We require attention to the interpretation of these commutation relations. 
In quantum mechanics, simultaneous observables are represented by commutative self-adjoint operators. 
If this interpretation is applied to our model, 
Eq.\ (\ref{eqn:commu_qxg}) means that 
the center of mass and relative coordinates are not simultaneously observable.
This is difficult to understand because both coordinates are known to be 
simultaneous observables in classical and quantum mechanics.
However, the quantum-mechanical relation between 
observables and commutativity is not directly applicable to the present toy model, because 
wave functions and simultaneous eigenstates are not defined at $t>0$. 
Thus we need further study to define the simultaneous measurement in this model.

\section{concluding remarks} \label{se:conclu}

We introduced the Poisson bracket operator using the operator derivative defined in quantum analysis.
This operator is an alternative quantum counterpart of the Poisson bracket in classical mechanics and there are at least three 
advantages compared to the commutator.
The operator derivative behaves as the standard derivative in the application to c-numbers  
and thus the Poisson bracket operator coincides with the Poisson bracket in the classical limit.
We further showed that the time differential equation of operators is
represented by using the Poisson bracket operator.
This is called the quantum canonical equation and agrees with the classical canonical equation in the classical limit.
This clear correspondence with classical mechanics is the first advantage of the introduction of the Poisson bracket operator.

In the standard applications of quantum mechanics, 
the Poisson bracket operator is expressed in terms of the commutator and then 
the quantum canonical equation is equivalent to the Heisenberg equation. 
At the same time, the quantum canonical equation is applicable to c-number canonical variables and then 
coincides with the classical canonical equation.
That is,  the Poisson bracket operator enables us to describe classical and quantum behaviors 
in a unified way and this is the second advantage.

The third advantage is that the quantum canonical equation is applicable to the system where the Heisenberg equation is not defined. 
As an example, we considered a toy model where a c-number and a q-number particles start to interact at an initial time.
The differential equations for the two particles satisfy the Ehrenfest theorem and 
are decomposed into the classical canonical equation for the c-number particle and the Heisenberg equation 
for the q-number particle in the non-interacting case. 
Moreover the conserved energy of the system is defined by the expectation value of the Hamiltonian operator.

If we identify the c-number and q-number particles of our toy model with, respectively, the classical and quantum particles, 
this model may be regarded as one of quantum-classical hybrids \cite{hall,elze,buric,koide_QCH}.
The description of our model is however incomplete and its consistency is still controversial.
For example, our model is described in the Heisenberg picture but the corresponding Schr\"{o}dinger picture is not defined. 
It is because the quantum canonical equations do not agree with the Heisenberg equations and the time evolution of operators is not represented by the unitary operator. 
As a result, the conservation of probability is not confirmed and 
simultaneous observables are not defined.

The Poisson bracket in classical mechanics is a canonical invariant, but the corresponding property in the Poisson bracket operator is 
not yet known. 
As a matter of fact, the property of the canonical transformation in quantum mechanics is not well understood. 
See, for example, Ref.\ \cite{bla2013} and references therein. 
One of the reasons for this difficulty is attributed to the fact that quantum mechanics is not necessarily defined 
for arbitrary generalized coordinate systems. 
For example, in polar coordinates, we need the operator representation 
of angle to describe the position of a particle.
However it is difficult to define the angle operator 
because there is no self-adjoint multiplicative operator which satisfies the periodicity 
and the canonical commutation relation simultaneously. 
This difficulty is the origin of the famous paradox in the angular uncertainty relation in the standard operator formulation of quantum mechanics. 
See also Refs.\ \cite{koide19,koide_ucr,water} and references therein.

We so far focus on quantized systems which are described by the commutator  (\ref{eqn:abcom}) and did not consider 
the fermionic system which is characterized by the anti-commutator.
The generalization of the present approach to the anti-commutator will be helpful to understand the classical correspondence of the anti-commutator.

By extending the procedure developed in this paper, the formulation of quantum mechanics will be described 
with the operator derivatives. 
The derivative has a geometrical meaning and thus its role is relatively easier to understand than 
that of the commutator.
Therefore, such a re-formulation will be useful to find the possible generalization of quantum mechanics.
For example, the definition of the standard derivative is affected by the curvature of geometry 
and thus a similar modification is expected to appear in the operator derivative. 
This perspective will be of assistance to develop quantum mechanics in curved geometry 
\cite{jensen,costa,ikegami,koide19}.

The semiclassical method has been used in various applications of quantum mechanics 
such as quantum chaos \cite{qchaos}, quantum-to-classical transition \cite{elze2}, 
semi-classical gravity \cite{semi-gra}, non-Hermitian generalization of quantum mechanics \cite{ptqm} 
and so on \cite{zurek}.
Such an approach will be helpful even to understand a macroscopic matter wave interference observed 
in extremely massive and complex molecules \cite{fllurene,kda}.
The formulation of the semiclassical theory is however not straightforward 
because of the singular behavior in the vanishing limit of $\hbar$ \cite{berry}. 
In the operator derivative (\ref{eqn:def_op_der}), quantum effects appear through the operator $\delta_{\widehat{A}}$. 
Therefore, 
by introducing the systematic expansion with respect to $\delta_{\widehat{A}}$ as discussed in Sec.\ \ref{sec:qc_eq}, 
it may be possible to develop a new approach which sheds new light on 
the semiclassical behavior of quantum mechanics.
The applications to these problems are left as future works.

\vspace*{1cm}
The author thanks T. Kodama for useful comments and acknowledges the financial support by CNPq (303468/2018-1).
A part of the work was developed under the project INCT-FNA Proc.\ No.\ 464898/2014-5.

\appendix

\section{Proof of Formula 1} \label{app:formula1}

This is proved by mathematical induction.  
It is easy to confirm that Eq.\ (\ref{eqn:mformula1}) is satisfied for $n=1$, 
\begin{eqnarray}
\int^1_0 d\lambda \, (\widehat{A}-\lambda \delta_{\widehat{A}} ) \widehat{B} 
&=& 
\frac{1}{2} (\widehat{A}\widehat{B} + \widehat{B}\widehat{A}) \, .
\end{eqnarray}

Suppose that Eq.\ (\ref{eqn:mformula1}) is satisfied for $n=L$ ($L \ge 1$). 
Then the left-hand side of Eq.\ (\ref{eqn:mformula1}) for $n=L+1$ is calculated as 
\begin{eqnarray}
\lefteqn{\int^1_0 d\lambda \, (\widehat{A}-\lambda \delta_{\widehat{A}} )^{L+1} \widehat{B} } && \nonumber \\
&=& 
\frac{(-1)^{L+1}}{L+2} (\delta_{\widehat{A}})^{L+1} \widehat{B} 
+ \frac{L+1}{L+2}
\sum^L_{m=0} 
{}_L C_m
(-1)^m  \left( \frac{1}{m+1} + \frac{1}{L+1-m}  \right) \widehat{A}^{L+1-m} (\delta_{\widehat{A}})^m \widehat{B} 
\nonumber \\
&=& 
\frac{L+1}{L+2} \widehat{A} \int^1_0 \ud \lambda \, (\widehat{A} - \lambda \delta_{\widehat{A}})^L \widehat{B}
 + \frac{1}{L+2}
\sum^{L+1}_{m=0} 
{}_{L+1} C_m
 (-1)^m  \widehat{A}^{L+1-m} (\delta_{\widehat{A}})^m \widehat{B}
\nonumber \\
&=& 
\frac{1}{L+2}\widehat{A}  (\widehat{A}^L \widehat{B} + \widehat{A}^{L-1}\widehat{B}\widehat{A} + \cdots + \widehat{B} \widehat{A}^L) 
+\frac{1}{L+2}
(\widehat{A}-\delta_{\widehat{A}})^{L+1} \widehat{B}
\nonumber \\
&=& 
\frac{1}{L+2}\widehat{A}  (\widehat{A}^L \widehat{B} + \widehat{A}^{L-1}\widehat{B}\widehat{A} + \cdots + \widehat{B} \widehat{A}^L) 
+\frac{1}{L+2}\widehat{B} \widehat{A}^{L+1} \, .
\end{eqnarray}
The last equality is the right-hand side of Eq.\ (\ref{eqn:mformula1}) for $n=L+1$. 
Therefore Eq.\ (\ref{eqn:mformula1}) is satisfied for any integer $n \ge 1$.
$\square$

\section{Proof of Formula 2} \label{app:formula2}

The operator $\widehat{B}^n \widehat{A}^m$  is reexpressed as
\begin{eqnarray}
\widehat{B}^n \widehat{A}^m 
&=& 
\widehat{B}^{n-1} (\delta_{\widehat{B}} \widehat{A} + \widehat{A} \widehat{B} ) \widehat{A}^{m-1} \nonumber \\
&=& 
(\delta_{\widehat{B}} \widehat{A}) \widehat{B}^{n-1}\widehat{A}^{m-1} + \widehat{B}^{n-1}\widehat{A} \widehat{B} \widehat{A}^{m-1} \nonumber \\
&=&
m (\delta_{\widehat{B}} \widehat{A}) \widehat{B}^{n-1}\widehat{A}^{m-1} 
+ \widehat{B}^{n-1} \widehat{A}^{m}  \widehat{B} \, .
\end{eqnarray}
Here we used $\delta_{\widehat{B}} \widehat{A} = -c$.
Applying this result to the operator $\widehat{B}^{n-1} \widehat{A}^{m} $ which appears in the second term on the right-hand side 
of the above equation, 
we find  
\begin{eqnarray}
\widehat{B}^n \widehat{A}^m 
&=&
m (\delta_{\widehat{B}} \widehat{A}) \widehat{B}^{n-1}\widehat{A}^{m-1} 
+ \left\{ m (\delta_{\widehat{B}} \widehat{A}) \widehat{B}^{n-2}\widehat{A}^{m-1} 
+ \widehat{B}^{n-2} \widehat{A}^{m}  \widehat{B} \right\}  \widehat{B}  \nonumber \\
&=&
m (\delta_{\widehat{B}} \widehat{A}) 
\left\{ 
\widehat{B}^{n-1}\widehat{A}^{m-1} 
+  \widehat{B}^{n-2}\widehat{A}^{m-1} \widehat{B}
\right\} 
+ \widehat{B}^{n-2} \widehat{A}^{m}  \widehat{B}^2  \nonumber \\
&=&
m (\delta_{\widehat{B}} \widehat{A}) 
\left\{ 
\widehat{B}^{n-1}\widehat{A}^{m-1} 
+  \widehat{B}^{n-2}\widehat{A}^{m-1} \widehat{B}
+ \cdots 
+ \widehat{B} \widehat{A}^{m-1} \widehat{B}^{n-2} 
+ \widehat{A}^{m-1} \widehat{B}^{n-1}
\right\} 
+ \widehat{A}^{m}  \widehat{B}^{n}  \nonumber \\
&=&
m n (\delta_{\widehat{B}} \widehat{A})  \int^1_0 \ud \lambda (\widehat{B} - \lambda \delta_{\widehat{B}})^{n-1} \widehat{A}^{m-1} + \widehat{A}^{m}  \widehat{B}^{n}  \, .
\end{eqnarray}
In the last line, the formula (\ref{eqn:mformula1}) was used. 
This is Eq.\ (\ref{eqn:mformula3}).
$\square$

\section{Proof of Formula 3} \label{app:formula3}

By the interchange between $(m,\widehat{A})$ and $(n,\widehat{B})$ in the formula (\ref{eqn:mformula3}), we obtain
\begin{eqnarray}
\widehat{A}^m \widehat{B}^n  - \widehat{B}^n  \widehat{A}^m  
&=& m n (\delta_{\widehat{A}} \widehat{B})  \int^1_0 \ud \lambda \, (\widehat{A} - \lambda \delta_{\widehat{A}})^{m-1} \widehat{B}^{n-1} 
\, .
\end{eqnarray}
Using this and Eq.\ (\ref{eqn:mformula3}) itself, we find 
\begin{eqnarray}
m n (\delta_{\widehat{B}} \widehat{A})  \int^1_0 \ud \lambda\, (\widehat{B} - \lambda \delta_{\widehat{B}})^{n-1} \widehat{A}^{m-1} 
= 
- m n (\delta_{\widehat{A}} \widehat{B})  \int^1_0 \ud \lambda\, (\widehat{A} - \lambda \delta_{\widehat{A}})^{m-1} \widehat{B}^{n-1} 
\, .
\end{eqnarray}
It is easy to obtain Eq.\ (\ref{eqn:mformula2}) from this because $\delta_{\widehat{B}} \widehat{A} = - \delta_{\widehat{A}} \widehat{B}$.
$\square$

\section{Ehrenfest theorem and non-interacting limit} \label{app:ehren}

The quantum canonical equations (\ref{eqn:model_eom}) satisfy the Ehrenfest theorem. 
Indeed we derive classical canonical equations from the classical Hamiltonian which is obtained by replacing q-numbers with the corresponding c-numbers in Eq.\ (\ref{eqn:totalh}).
The structures of these classical canonical equations coincide with our quantum canonical equations (\ref{eqn:model_eom}) 
when the non-commutativity of q-numbers is ignored.

Below Eq.\ (\ref{eqn:sys_h2}), we discussed the desirable property 
which should be satisfied in the case of no interaction $\alpha = 0$.
This property is hold in our model.
Then the solution of Eqs.\ (\ref{eqn:eq_com}) and (\ref{eqn:eq_qpq})
are reduced to 
\begin{eqnarray}
\begin{split}
x(t) 
&= \widehat{X}_C (t) + \frac{\uM}{\um + \uM} \widehat{q}(t) 
= x_0 + \frac{p_0 }{\um}t \, ,\\
p(t)
&= 
\widehat{p}_q (t) + \frac{\um}{\um + \uM} \widehat{P}_C (t)
=
p_0 \, ,\\
\widehat{X}(t)
&= 
\widehat{X}_C (t) - \frac{\uM}{\um + \uM} \widehat{q}(t)
=
\widehat{X}_0 + \frac{\widehat{P}_0}{\uM}t \, ,\\
\widehat{P}(t)
&= 
-\widehat{p}_q (t) + \frac{\uM}{\um + \uM} \widehat{P}_C (t)
=
\widehat{P}(t) 
 \, .
\end{split}
\end{eqnarray}
It is easy to confirm that these are the solutions of  the classical canonical equations (\ref{eqn:noint1}) and 
the Heisenberg equations (\ref{eqn:noint2}).
Moreover, the commutation relations of these canonical variables satisfy Eq.\ (\ref{eqn:commu_noint}) and thus 
the canonical variables $({x}(t), {p}(t))$ recover commutativity as we expected 
in the case of no interaction.

\section{conservation laws} \label{app:conserve}

In this Appendix, the conservation laws of the model in Sec.\ \ref{sec:app} are discussed.
The total momentum conservation is easily seen from the behavior of $\widehat{P}_C (t)$ shown in 
Eq.\ (\ref{eqn:eq_com}).

The total energy of this system is defined by the expectation value of the total Hamiltonian operator, 
$\langle \Psi | H | \Psi \rangle$,
where $\Psi$ is the initial wave function.
The time evolution of the Hamiltonian operator (\ref{eqn:totalh}) is determined by  
\begin{eqnarray}
\frac{\ud}{\ud t} H 
= 
\left\{ H, H \right\}_{(x(t),p(t);\widehat{X}(t),\widehat{P}(t))} \, .
\label{eqn:dhdt}
\end{eqnarray}
Thus, to conserve the energy, $\left\{ H, H \right\}_{(x(t),p(t);\widehat{X}(t),\widehat{P}(t))}$ should vanish.
This is however not trivial because the Poisson bracket operator is not equivalent to the commutator in the present toy model as is shown later in Sec.\ \ref{sec:commu}.

To show the energy conservation, we need to calculate directly the Poisson bracket operators using the Hamiltonian (\ref{eqn:totalh}). 
We then find 
\begin{eqnarray}
\left\{ H, H \right\}_{(x(t),p(t))} 
= 
\left\{ H, H \right\}_{(\widehat{X}(t),\widehat{P}(t))} = 0 \, .
\end{eqnarray}
In this calculation, we used 
\begin{eqnarray}
\frac{\partial H}{\partial x(t)} \left\{ \frac{\partial H}{\partial p(t)} \right\}
&=& 
 \frac{\alpha}{2} \left(\int^1_0 \ud \lambda
2( x(t) - \lambda \delta_{\widehat{x}(t)}) \frac{p(t)}{\um} - \widehat{X}(t) \frac{p(t)}{\um}
- \frac{p(t)}{\um}\widehat{X}(t) 
\right) \nonumber \\
&=& 
\frac{\alpha}{2\um} \left\{ p(t)  (x(t) - \widehat{X}(t))  + (x(t) - \widehat{X}(t)) p(t) \right\} \, .
\end{eqnarray}
Therefore the total energy of this model is conserved.

\end{document}